\documentclass{aa}
\usepackage{tabularx}
\usepackage{color}
\usepackage{multirow}
\usepackage{txfonts}
\usepackage{tabularx}
\usepackage{bm}
\usepackage{epstopdf}
\usepackage{subfigure}

\usepackage{amsmath}
\usepackage{ulem}

\usepackage{comment}

\begin{document}
\title{CHANG-ES XV: Large-scale magnetic field reversals in the radio halo of NGC~4631 \thanks{Based on observations
with the Karl G. Jansky Very Large Array (VLA) operated by the NRAO. The NRAO is a facility of the National Science Foundation operated under agreement by Associated Universities, Inc.}}
\author{Silvia Carolina Mora-Partiarroyo\inst{1} \thanks{\email{silvia.carolina.mora@gmail.com}} \and
Marita Krause\inst{1} \thanks{\email{mkrause@mpifr-bonn.mpg.de}}
\and
Aritra Basu\inst{2}
 \and
 Rainer Beck\inst{1}
 \and
 Theresa Wiegert\inst{3}
 \and
 Judith Irwin\inst{3}
 \and 
 Richard Henriksen\inst{3}
 \and 
 Yelena Stein\inst{4,5}
 \and
 Carlos J. Vargas\inst{13}
 \and
 Volker Heesen\inst{6}
 \and
 Ren{\'e} A. M. Walterbos\inst{7}
 \and
 Richard J. Rand\inst{8}
 \and
  George Heald\inst{9}
 \and
 Jiangtao Li\inst{10}
 \and
 Patrick Kamieneski\inst{11}
 \and
 Jayanne English\inst{12}
}
\institute{Max-Planck-Institut f\"ur Radioastronomie, Auf dem H\"ugel 69, 53121 Bonn, Germany
\and
Fakult{\"a}t f{\"u}r Physik, Universit{\"a}t Bielefeld, Postfach 100131, 33501 Bielefeld, Germany
\and
Department of Physics, Engineering Physics, \& Astronomy, Queen's University, Kingston, ON, K7L 3N6, Canada
\and
Observatoire Astronomique de Strasbourg, Universit{\'e} de Strasbourg, CNRS, UMR 7550, 11 rue de l’Universit{\'e},
67000 Strasbourg, France
\and
Astronomisches Institut (AIRUB), Ruhr-Universit{\"a}t Bochum, Universit{\"a}tsstrasse 150, 44801 Bochum, Germany
\and
Hamburger Sternwarte, Universit{\"a}t Hamburg, Gojenbergsweg 112, 21029 Hamburg, Germany
\and
Department of Astronomy, New Mexico State University, Las Cruces, NM 88001, U.S.A.
\and
Department of Physics and Astronomy, University of New Mexico, 1919 Lomas Blvd. NE, Albuquerque, NM 87131, U.S.A.
\and
CSIRO Astronomy and Space Science, PO Box 1130, Bentley, WA 6102, Australia
\and
Dept. of Astronomy, University of Michigan, 311 West Hall, 1085 S. University Ave., Ann Arbor, MI 48109, U.S.A.
\and
Dept. of Astronomy, University of Massachusetts, 710 North Pleasant St., Amherst, MA, 01003, U.S.A.
\and Department of Physics and Astronomy, University of Manitoba, Winnipeg, Manitoba, R3T 2N2, Canada
\and
Department of Astronomy and Steward Observatory, University of Arizona, 933 N Cherry Ave, Tucson, AZ 85719 U.S.A.}
\date{Received 04.11.2018 / Accepted 29.07.2019}

\abstract{} 
{NGC~4631 is an interacting galaxy which exhibits one of the largest gaseous halos observed among edge-on galaxies. We aim to examine the synchrotron and polarization properties of its disk and halo emission with new radio continuum data. }
{Radio continuum observations of NGC~4631 were performed with the Karl G. Jansky Very Large Array at C-band (5.99~GHz) in the C \& D array configurations, and at L-band (1.57~GHz) in the B, C, \& D array configurations. The Rotation Measure Synthesis algorithm was utilized to derive the polarization properties.}
{We detected linearly polarized emission at C-band and L-band. The magnetic field in the halo is characterized by strong vertical components above and below the central region of the galaxy. The magnetic field in the disk is only clearly seen in the eastern side of NGC~4631, where it is parallel to the plane of the major axis of the galaxy. We detected for the first time a large-scale, smooth Faraday depth pattern in a halo of an external spiral galaxy, which implies the existence of a regular (coherent) magnetic field. A quasi-periodic pattern in Faraday depth with field reversals was found in the northern halo of the galaxy. }
{The field reversals in the northern halo of NGC~4631, together with the observed polarization angles, indicate giant magnetic ropes (GMRs) with alternating directions. To our knowledge, this is the first time such reversals are observed in an external galaxy.}

\keywords{Galaxies: individual: NGC~4631 -- galaxies: halos -- galaxies: magnetic fields -- galaxies: interactions -- galaxies: spiral -- radio continuum: galaxies}

\titlerunning{CHANG-ES XV: Large-scale magnetic fields in the halo of NGC 4631}
\authorrunning{Mora-Partiarroyo et al.} 
\maketitle

\section{Introduction}
\label{intro}

The galaxy NGC~4631 is known for hosting dominant vertical magnetic fields in its radio halo\footnote{In this paper we use the terms ``halo'' and ``thick disk'' synonymously.} \citep{Hummel1991} which were sometimes claimed to even cross the plane of the disk \citep{Golla1994}. More recent studies, however, found that the disk itself contains a plane-parallel magnetic field \citep{Krause2009, Mora2013}, similar to other galaxies observed face-on and edge-on. The observed magnetic field orientation in the halo shows an X-shaped halo field with strong vertical components above and below the central region, as typically found in edge-on galaxies \citep{Krause2011}. Up to now, it is still unclear whether the magnetic field in the halo is indeed regular (coherent) or anisotropic (e.g., elongated filaments or loops). Only measurements of the Faraday depth (Faraday rotation measure) can distinguish between the two magnetic field configurations: while the Faraday depth cancels along the line-of-sight (LOS) and/or frequently changes sign in the presence of an anisotropic field, it adds up along the LOS or varies smoothly over several beams in the presence of a regular magnetic field. Furthermore, the sign of the Faraday depth reveals the direction of the magnetic field component parallel to the LOS, which is needed to identify -- together with the orientation of the magnetic field --  a large-scale magnetic field pattern in the halo. 

In this paper we present polarization observations of NGC~4631 with the Karl. G. Jansky Very Large Array (VLA) at C-band (5.99~GHz) and L-band (1.57~GHz).
The observations and data reduction are described in \cite{Mora-Partiarroyo2019_A}, together with an analysis of the total power and synchotron emission. There was no need for a combination with single-dish Effelsberg data because our VLA images in Stokes $Q$ and $U$, smoothed to the resolution of the Effelsberg telescope, do not suffer from missing large-scale emission. Results for polarized intensities (PI) and Faraday depth of the galaxy
are presented in Sect.~\ref{section_polarization}, followed by a discussion in Sect.~\ref{Discussion}, and conclusions in Sect.~\ref{Conclusions}.
The parameters of NGC~4631 assumed in this study are presented in Table~1 of \cite{Mora-Partiarroyo2019_A}. In all radio maps presented in this paper, the beam area is shown as a filled circle in the left-hand corner of each image.

\section{Data reduction}
%\label{Observations}
\label{VLA_observations}

Radio polarimetric observations of NGC~4631 were performed with the Karl G. Jansky Very Large Array (VLA) during its commissioning phase through the project \textsl{Continuum HAlos in Nearby Galaxies -- an EVLA Survey (CHANG-ES)}. 
The data were taken at C-band (5.0 -- 7.0 GHz) in C and D array configuration, and at L-band (1.25 -- 1.50 and 1.65 -- 1.90~GHz) in the B, C, and D array configurations. We refer to Sect. 2.1 in \cite{Mora-Partiarroyo2019_A} for details of the observations and standard data reduction.   

To apply Rotation Measure Synthesis (RM-synthesis, \citealt{Brentjens2005}) at C-band, the self-calibrated visibilities averaged over $\sim$125~MHz were imaged in Stokes \textsl{Q} and \textsl{U} employing the multi-scale deconvolution available in the Common Astronomy Software Applications package (CASA; \citealt{2007CASA}). Both pointings at each of the 16 central frequencies were smoothed to the lowest angular resolution (i.e., the size of the synthesized beam at the lowest central frequency), and were PB-corrected and mosaicked during the cleaning routine. The 16 mosaicked images were thereafter combined into a cube. 
For the L-band data, a similar procedure was followed, but in this case the self-calibrated visibilities were imaged per 0.5-MHz wide channel, since the electric field vector rotates more rapidly at lower frequencies and therefore, bandwidth depolarization is more severe. We applied the RM-clean algorithm \citep{Heald2009} to clean the effects of the RM Spread Function (RMSF) in the Faraday depth cubes down to $5\sigma$ level of the noise in Stokes \textsl{Q} and \textsl{U}. 
The resolution and maximum scale in Faraday space at each frequency band are presented in Table \ref{RM_parameters}.

\begin{table}[h]
\centering
\caption[]{Faraday resolution ($\delta \phi$) and maximum scale in Faraday space ($\rm{scale_{max}}$) at each frequency band obtained with RM-synthesis.}
\label{RM_parameters}
\begin{tabularx}{0.8\columnwidth}{c*3{>{\centering\let\\=\tabularnewline}X}}
\hline \hline
Frequency band & ${\rm \delta \phi}$ & $ \rm{scale_{max}}$ \\
& \small{[$\rm{ rad/m^{2}}$]} & \small{[${\rm{ rad /m^{2}}}$]} \\ \hline  
C-band & 2065 & 1691 \\ 
L-band & 107 & 126  \\
\hline
\end{tabularx}
\end{table}

The polarized intensity and polarization angle at any given pixel of the map were determined using the Stokes \textsl{Q} and \textsl{U} values at the peak in the Faraday spectrum. The data were corrected for the Ricean bias in polarized intensity as described in \cite{Adebahr2017}. The position of this peak defines the Faraday depth value (rotation measure). The polarization angles were corrected with the corresponding Faraday depth to obtain the intrinsic magnetic field orientation and were determined only where Stokes \textsl{Q} and \textsl{U} exceeded a 5$\sigma$ threshold.

\section{Results}
\subsection{Polarized intensities}
\label{section_polarization}

\begin{figure*}[]
\centering
\includegraphics[trim = 10mm 150mm 20mm 15mm, clip, width=1.\textwidth]{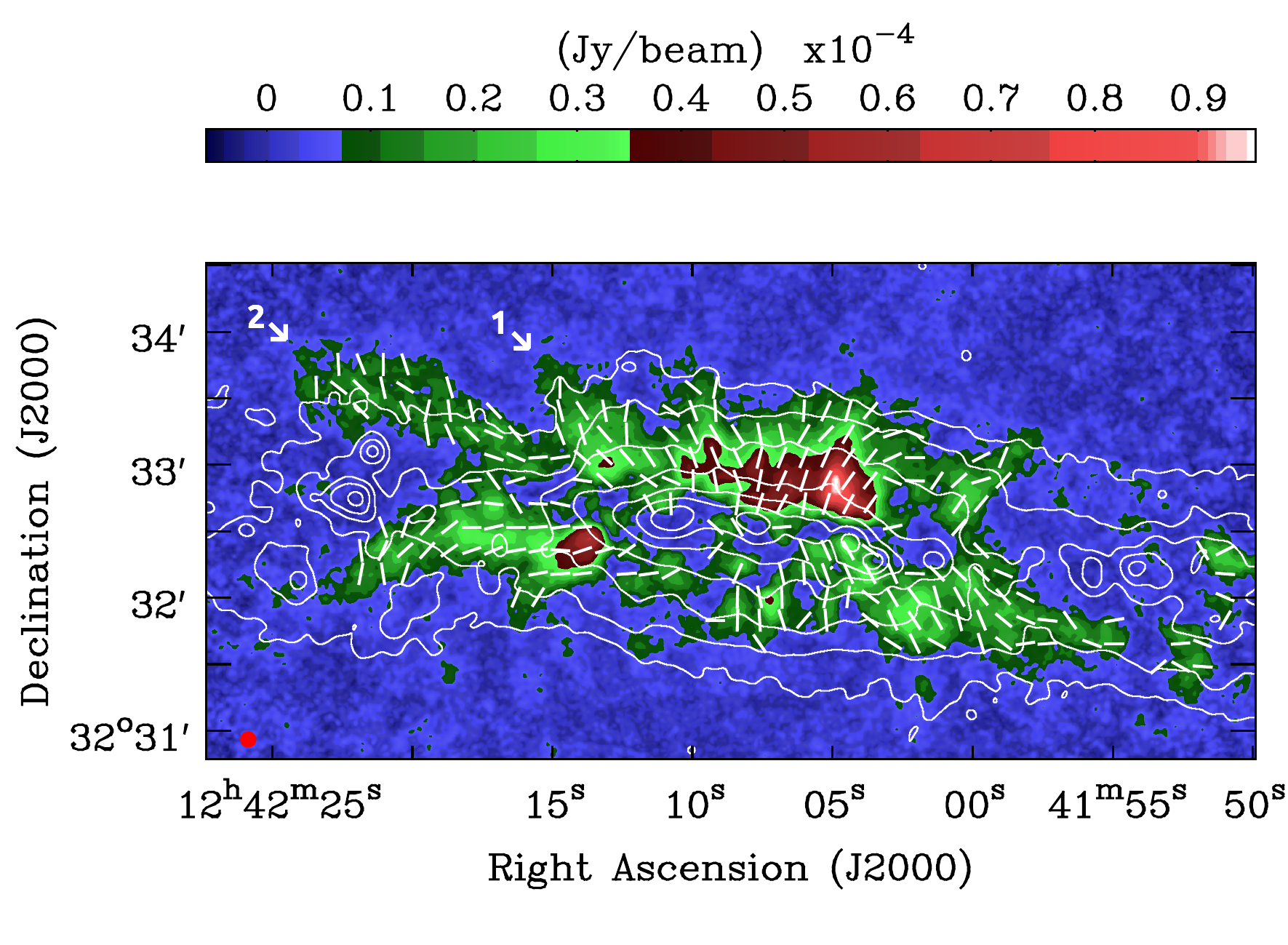}
\caption[]{Colorscale of the polarized emission of NGC~4631 at C-band (C+D configurations combined) with the intrinsic magnetic field orientation (line segments of equal
length) at C-band obtained with RM-synthesis. Polarization angles were calculated at pixels where the polarized intensity is larger than 5$\sigma$ (where $\sigma$ is the rms noise in Stokes \textsl{Q} \& \textsl{U}). Contours correspond to the 5.99~GHz total power emission (VLA + Effelsberg), refer to \cite{Mora-Partiarroyo2019_A} for details concerning the total intensity emission. All data plotted have an angular resolution of 7$''$~FWHM. Contour levels are at $18~\rm{\mu Jy/beam}\times(3, 6, 12, 24, 48, 96, 192, 384)$. The polarized spurs in the north-eastern quadrant have been labeled 1 and 2.}
\label{Cband_POLI_7arcsec}
\end{figure*}

\begin{figure}[h!]
\centering
\includegraphics[trim = 20mm 120mm 10mm 18mm, clip, width=1\columnwidth]{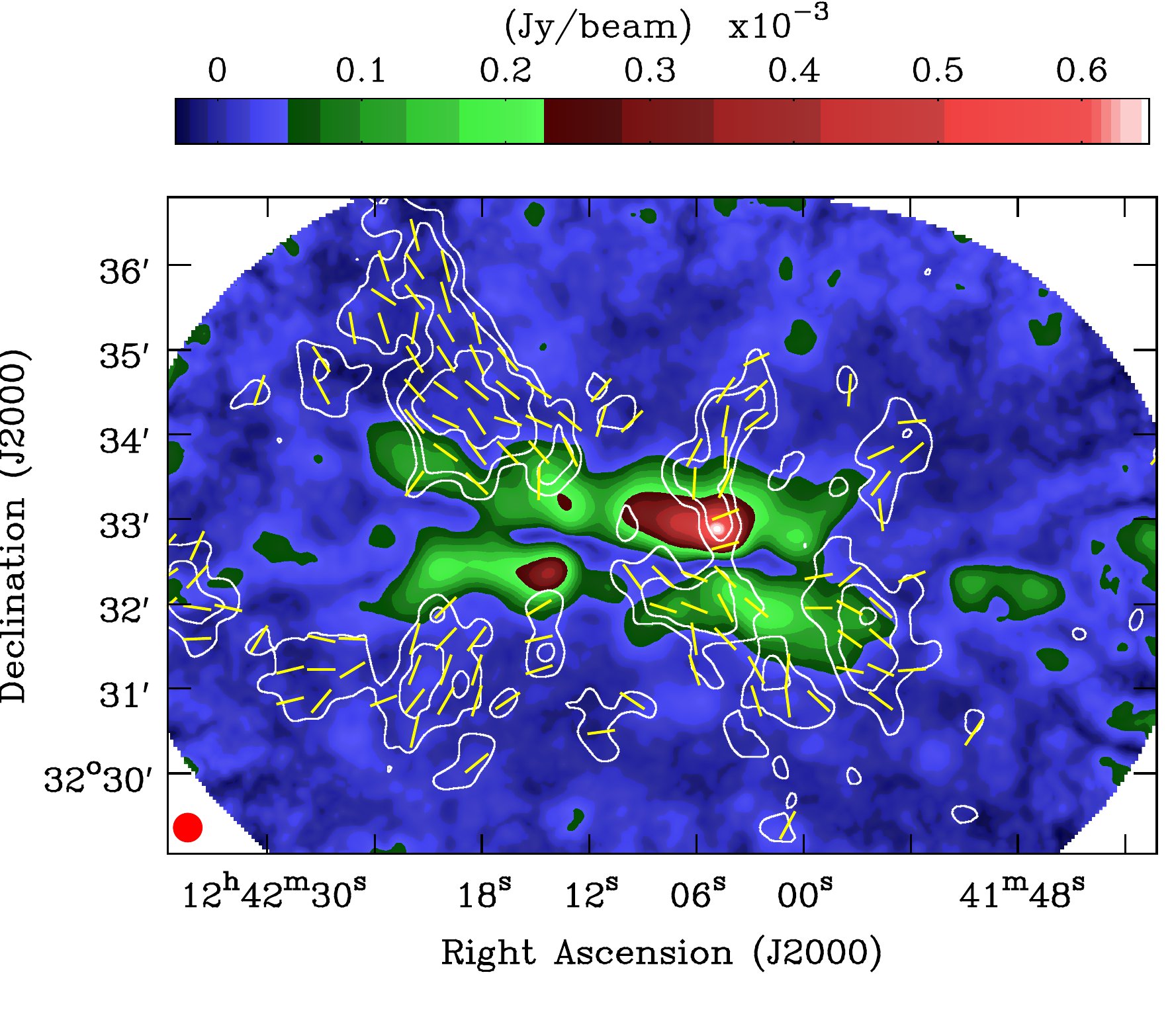}
\caption[]{Colorscale of the polarized emission of NGC~4631 at C-band (VLA D-configuration) together with the magnetic field orientation (line segments of equal length) obtained with the L-band (VLA C\&D-configuration) data. Polarization angles were calculated at pixels where the polarized intensity is larger than 5$\sigma$. All data plotted have an angular resolution of $20\farcs5$~FWHM. Contours correspond to the polarized emission at L-band (VLA C\&D-configuration) and are at $35~\rm{\mu Jy/beam}\times(3, 4.5, 7)$. }
\label{Cband_POLI_20.5arcsec}
\end{figure}

\begin{figure}[h!]
\centering
\includegraphics[trim = 30mm 140mm 20mm 18mm, clip, width=1\columnwidth]{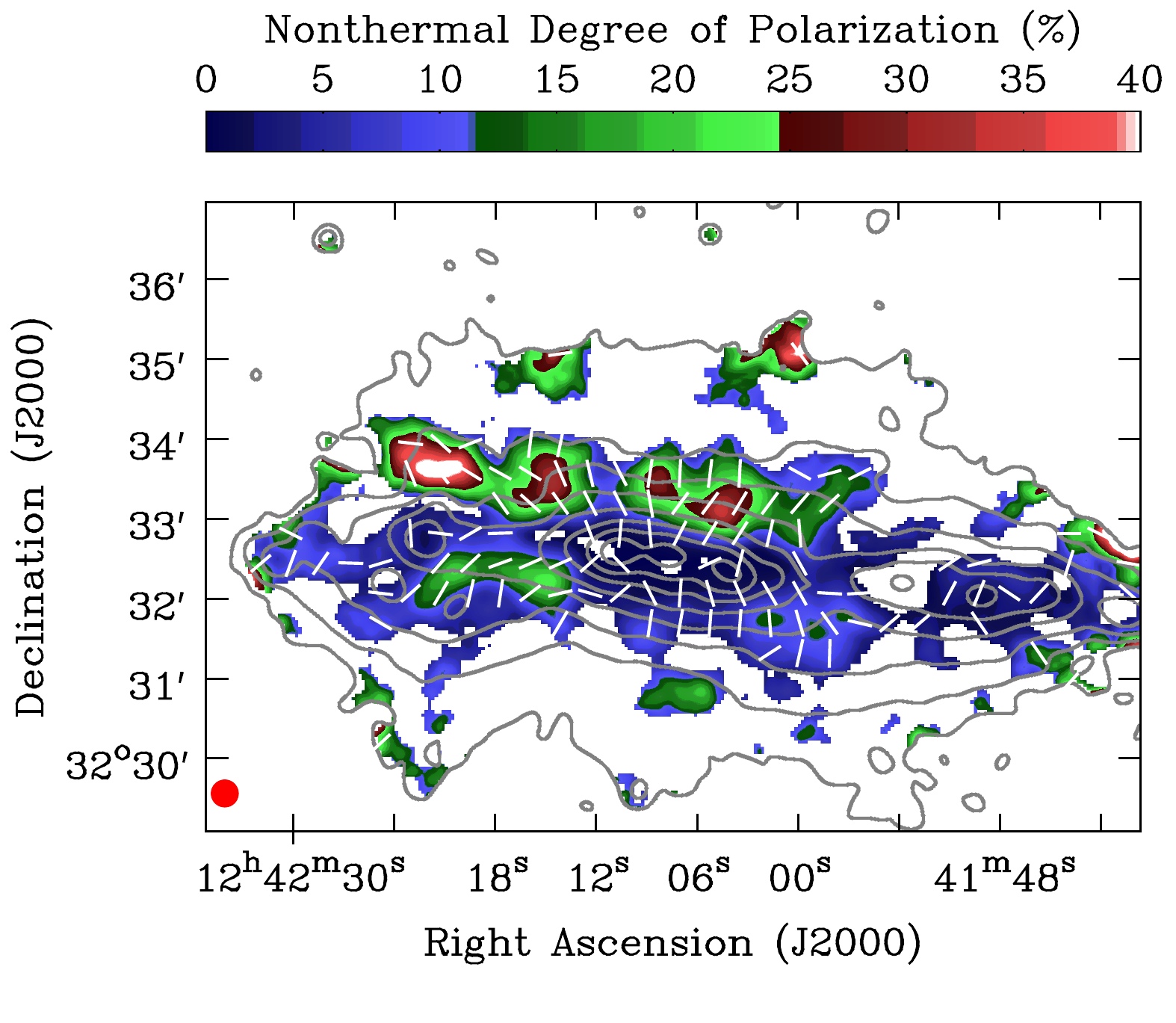}
\caption[]{Colorscale of the nonthermal degree of polarization at C-band with the intrinsic magnetic field orientation (line segments of equal length) at C-band. Values were calculated at pixels where the polarized intensity is larger than 2$\sigma$ and the synchrotron intensity is larger than 3$\sigma$ \citep{Mora-Partiarroyo2019_A}. C-band polarization angles (VLA D-configuration) were calculated at pixels where the polarized intensity is larger than 5$\sigma$. All data plotted have an angular resolution of $20\farcs5$~FWHM. Grey contours correspond to the 5.99~GHz total power emission (VLA + Effelsberg) are at $45~\rm{\mu Jy/beam}\times(3, 6, 12, 24, 48, 96, 192, 384)$. } 
\label{Cband_POLIDEG_20.5arcsec}
\end{figure}

The polarized emission obtained at C-band is shown in Figures~\ref{Cband_POLI_7arcsec} and \ref{Cband_POLI_20.5arcsec} at angular resolutions of 7$''$ and 20$\farcs$5, respectively. The distribution of the nonthermal degree of polarization (PI/I$_{\rm syn}$) is presented in Figure~\ref{Cband_POLIDEG_20.5arcsec}. This is the first time the polarized emission of NGC~4631 has been observed at such high resolution.

\begin{figure}[h!]
\centering
\includegraphics[trim = 40mm 120mm 15mm 15mm, clip, width=1\columnwidth]{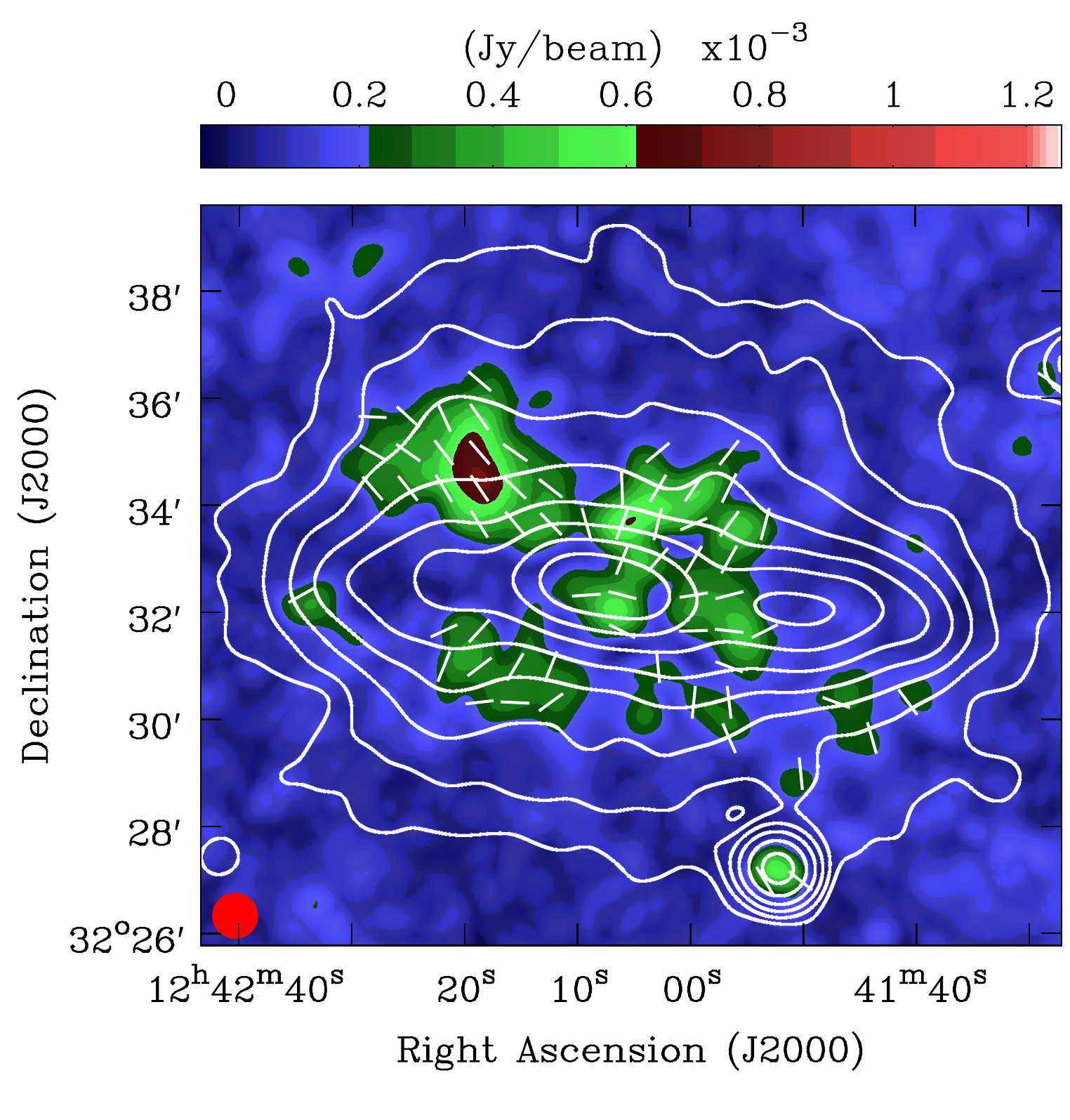}
\caption[]{Colorscale of the polarized emission of NGC~4631 at L-band (VLA D-configuration), together with the magnetic field orientation (line segments of equal length) at L-band obtained with RM-synthesis. Polarization angles were calculated at pixels where the polarized intensity is larger than 5$\sigma$. Contours correspond to the 1.57~GHz total power emission (VLA + Effelsberg; \citealt{Mora-Partiarroyo2019_A}). All data plotted have an angular resolution of 51$''$~FWHM. Contour levels are at $200~\rm{\mu Jy/beam}\times(3, 6, 12, 24, 48, 96, 192, 384)$.}
\label{Lband_POLI_51arcsec}
\end{figure}

The linearly polarized emission L-band is presented in white contours in Figure~\ref{Cband_POLI_20.5arcsec} at a resolution of $20\farcs5$ and in color scale in Figure~\ref{Lband_POLI_51arcsec} at a resolution of 51$''$. 
Due to the higher sensitivity of our observations and the application of RM-synthesis, we recovered more polarized emission in the halo at both wavelengths than in previous VLA observations by \citet{Irwin2012_I}. Although the Faraday depolarization decreases with increasing frequency, the sensitivity in C-band is generally lower than at L-band, so that polarized emission in C-band is not detected towards the outer halo.

\subsection{Faraday depth distribution and polarization angles}
\label{FDdistribution}

\begin{figure*}[h]
\centering
\includegraphics[trim = 10mm 145mm 25mm 15mm, clip, width=1.80\columnwidth]{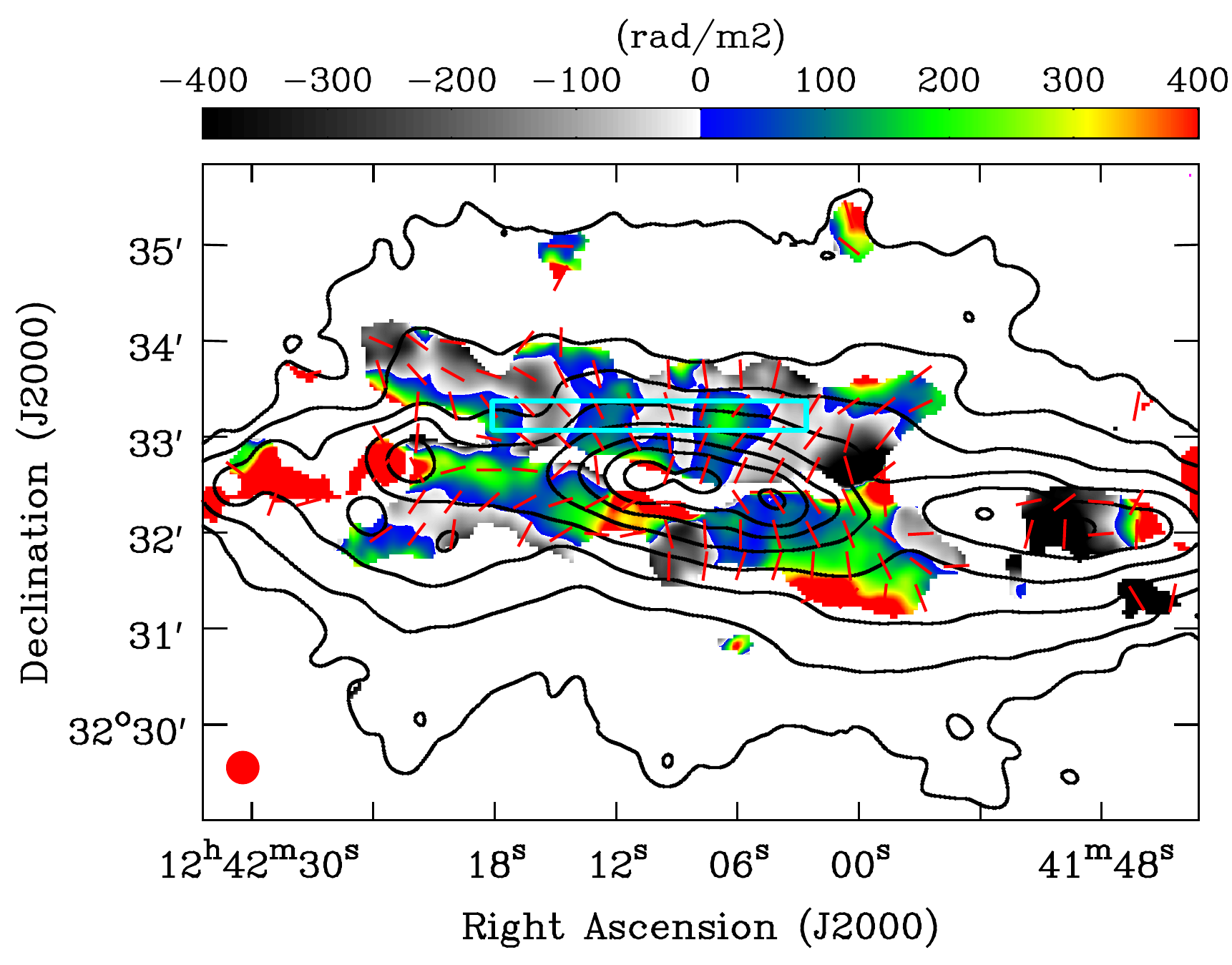}
\caption[]{Colorscale of the Faraday depth obtained from C-band (VLA D-configuration)  data, together with the intrinsic magnetic field orientation (line segments of equal length) at C-band. Faraday depth and polarization angles were calculated at pixels where the C-band polarized intensity is larger than 5$\sigma$. All data plotted have an angular resolution of $20\farcs5$~FWHM. Black contours corresponding to the 5.99~GHz total power emission (VLA + Effelsberg) are at $45~\rm{\mu Jy/beam} \times(3, 6, 12, 24, 48, 96, 192, 384)$, refer to \cite{Mora-Partiarroyo2019_A} for details concerning the total intensity emission. The blue rectangular box indicates the region from which the Faraday depth profile shown in Fig.~\ref{profile} was extracted.}
\label{RM_Cbad_20.5arcsec}
\end{figure*}

The position of the peak in polarized intensity at a given pixel along the Faraday depth axis (Faraday spectrum) defines the corresponding Faraday depth at the pixel.
At C-band, the Faraday depths range between $-400$ and $400~\rm{rad / m^{2}}$. At L-band, they range between $-40$ and $40~\rm{rad / m^{2}}$, since observations at this frequency band are limited by the maximum scale in Faraday space of $126~\rm{rad / m^{2}}$. Missing spacings in wavelength coverage create artifacts affecting the detected Faraday depth, as demonstrated by \cite{Schnitzeler2015_II}, which is the case for our L-band observations. In addition, we are probably probing different regions along the LOS at C-band and at L-band due to the strong wavelength dependence of Faraday depolarization effects \citep{Burn, Sokoloff}. Only at C-band, can we expect to trace a large LOS through the galaxy and perhaps even its entire extent.

A map of the Faraday depth, at an angular resolution of $20\farcs5$, computed using the C-band data, is presented in Figure~\ref{RM_Cbad_20.5arcsec}. In this figure, the maximum error in Faraday depth is about 80$~\rm{rad / m^{2}}$ and decreases to about 30$~\rm{rad / m^{2}}$ in areas of high polarized intensity (refer to the color scale of PI in Fig.~\ref{Cband_POLI_20.5arcsec}). The Galactic foreground component is negligible in the direction of the sky in which NGC~4631 is located. \citet{Heald2009} obtained a Galactic foreground of $-4\pm3~\rm{rad / m^{2}}$ and \citet{Oppermann2012} estimated a value of $-0.3\pm2.7~\rm{rad / m^{2}}$.

For better visual presentation, we plotted in Figure~\ref{RM_Cbad_20.5arcsec} positive Faraday depth values in a color scale, while negative Faraday depth values are presented in a gray scale. Even at C-band, the midplane of the galaxy is completely depolarized (see Figs.~\ref{Cband_POLI_20.5arcsec} and \ref{Cband_POLIDEG_20.5arcsec}). The halo of NGC~4631 can be traced in Faraday depth up to a height of about 2--3~kpc (with an angular resolution that corresponds to about 750~pc), offering the unprecedented possibility to detect systematic patterns in the Faraday depth map. From Figure~\ref{RM_Cbad_20.5arcsec}, we identify large-scale regions (significantly larger than the beam size) in the northern and southern halo of NGC~4631, that show positive Faraday depth values and others that show negative values, with a smooth transition between them. This smooth, large-scale Faraday depth pattern is the first direct observational evidence for the existence of a regular (coherent) magnetic field in the halo of a spiral galaxy. This will be further discussed in Section~\ref{Discussion}.

The intrinsic magnetic field orientation in the plane of the sky (i.e., corrected for the Faraday rotation) derived from C-band data is shown in Figures~\ref{Cband_POLI_7arcsec}, \ref{Cband_POLIDEG_20.5arcsec}, and \ref{RM_Cbad_20.5arcsec}. The magnetic field orientation derived from the L-band data is presented in Figures~\ref{Cband_POLI_20.5arcsec} and \ref{Lband_POLI_51arcsec}. The line segments which represent the magnetic field orientation have an equal length in each map; this length was arbitrarily selected for better visualization. At both wavebands, we can distinguish that the overall magnetic field structure in the halo of NGC~4631 is X-shaped. At L-band, this is more evident at lower resolution (Fig.~\ref{Lband_POLI_51arcsec}). In addition, at C-band, the magnetic field in the halo is also characterized by strong vertical components above and below the central region of the galaxy, which are almost perpendicular to the major axis of NGC~4631. The vertical components above the central area of the galaxy had already been observed \citep{Golla1994}, but the vertical components below the major axis had never been detected at high resolution.

Along the central part of the disk observed at L-band, the magnetic field is oriented parallel to the midplane of the galaxy over several synthesized beams, in agreement with \cite{Mora2013}. 
This field parallel to the major axis of the galaxy is probably caused by a layer of emission observed on the front side of the disk, since the central region is affected considerably by Faraday depolarization at L-band.

At C-band, the magnetic field in the disk is only clearly seen in the eastern side of NGC~4631, where it is parallel to the plane of the major axis of the galaxy. Figure~\ref{Cband_POLIDEG_20.5arcsec} demonstrates that the disk is highly depolarized at C-band. In the central region of the galaxy we find an average nonthermal degree of polarization of $\simeq$1.5\%, while the eastern and western sides of the disk have a higher polarization fraction of about 6\% and 3\%, respectively (refer to \cite{Mora-Partiarroyo2019_A} for details concerning the synchrotron emission). The high spectral resolution of our data and application of the RM-synthesis algorithm are able to reduce bandwidth depolarization and differential Faraday rotation depolarization, but it cannot account for Faraday dispersion or beam depolarization. As demonstrated by \cite{Mora2013}, observations at 8.35~GHz are less affected by Faraday depolarization effects and therefore, the magnetic field along the midplane is more clearly visible at higher frequencies.

\subsection{Polarized spurs}
\label{PI_spurs_and_Bfield}

In contrast to what is observed at C-band, the highest polarized intensities and polarization fractions at L-band are found in the north-eastern quadrant of the galaxy. This is attributed to a spur-like feature observed at L-band that extends towards the north-eastern halo of NGC~4631. This spur is clearly visible in Figure~\ref{Cband_POLI_20.5arcsec} and it has a nonthermal degree of polarization of up to 44\% in the outer halo. This feature coincides with the NE-radio spur observed in total power (see Sect. 3.1.1 in  \citealt{Mora-Partiarroyo2019_A}). Furthermore, at the base of this spur (in the inner north-eastern halo) one can distinguish spur-like counterparts at C-band (spur 1 in Fig.~\ref{Cband_POLI_7arcsec} and see also Fig.~\ref{Cband_POLI_20.5arcsec}) and at 8.35~GHz (see Fig.~8 of \citealt{Mora2013}) which originate in the eastern end of the central region. The C-band counterpart has a nonthermal polarization fraction up to 32\%. At higher angular resolution, the polarization observed at C-band (Fig.~\ref{Cband_POLI_7arcsec}) shows that the spur is connected to the total radio emission coinciding with the H{\sc ii} region CM~67.

We wish to emphasize a spur-like feature of polarized emission located in the north-eastern quadrant of the galaxy at C-band (spur 2 in Fig.~\ref{Cband_POLI_7arcsec}) and at 8.35~GHz (Fig.~8 of \citealt{Mora2013}).  
It is also anchored in the eastern part of the central area of the galaxy and extends to the north-east. It has a nonthermal polarization fraction of up to 44\% at C-band. The L-band polarized emission displays structure at $\alpha_{2000}=12^\mathrm{h}42^\mathrm{m}27^\mathrm{s}; \delta_{2000}= 32^{\circ}34'30''$ which may be an extension of this spur (see Fig.~\ref{Cband_POLI_20.5arcsec}) having a nonthermal degree of polarization of up to 45\%. This high degree of polarization implies that differential Faraday depolarization is negligible and that the magnetic field along the spur is highly ordered. This could indicate an anisotropic magnetic field due to compression or stretching, or it may point to a regular magnetic field structure. This will be further discussed in Sect.~\ref{Discussion}.

There is another spur-like feature in the south-eastern quadrant of the galaxy; at C-band it emerges from the eastern side of the central region, extends towards the east and then bends south-east. It has a nonthermal degree of polarization of up to 23\% at C-band. A counterpart is clearly observable at 8.35~GHz \citep{Mora2013}, which also extends to the east and then bends south-east. Figure~\ref{Cband_POLI_20.5arcsec} indicates that at L-band this spur might extend farther out into the south-eastern halo, down to a declination of about $\delta_{2000}= 32^{\circ}30'$ with a nonthermal degree of polarization of up to 21\%. It appears to be connected to the SE-radio spur observed in total power (refer to Sect. 3.1.1 in  \citealt{Mora-Partiarroyo2019_A}). The orientation of the magnetic field of this polarized spur in the south-eastern halo seems to be aligned with the H{\sc i} spur 2 (see Fig.~3 in  \citealt{Mora-Partiarroyo2019_A}). Since this is also parallel to the X-shaped halo field, it is difficult to distinguish whether a part of the halo magnetic field is related to the H{\sc i} spur 2. Therefore, we cannot determine whether these features are actually related and if the magnetic field lines are coupled to the outflowing gas.

\subsection{Ratio between the size of turbulent cells and their filling factor}

Small-scale variations of magnetic field strength or orientation and/or variations in thermal electron density in a synchrotron emitting and Faraday rotating media cause depolarization that is called internal Faraday dispersion ($\rm DP_{IFD}$). This effect depends on the dispersion of intrinsic rotation measure within the volume of the telescope beam ($\sigma_{\rm RM}$) and on the wavelength of the observations. At longer wavelengths (i.e., lower frequencies), the observed Faraday depolarization is expected to be dominated by internal Faraday dispersion effects due to turbulence in the interstellar medium. Therefore, we derived the RM dispersion ($\sigma_{\rm RM}$) considering the ratio of the polarization degree ($p$) of the nonthermal emission at two frequencies:
\begin{equation}
\label{my_formula}
\rm{DP_{IFD}} = \frac{p_1}{p_2} = \left(\frac{1- \exp(-2\lambda_1^{4}\sigma_{RM}^{2})}{2\lambda_1^{4}\sigma_{RM}^{2}}\right) \left(\frac{2\lambda_2^{4}\sigma_{RM}^{2}}{1- \exp(-2\lambda_2^{4}\sigma_{RM}^{2})}\right)
\end{equation}
Here, the subscripts `1' and `2' refer to higher and lower
frequencies, respectively.
The ratio between the size of turbulent magnetic field cells ($d$) and the volume filling factor ($f$) of the cells, $d/f$, then follows from the RM dispersion \citep{Arshakian2011}: $\rm{d/f \cong \sigma_{RM}^{2}/([0.81 \; \langle n_e\rangle \; \langle B_{t,\|}\rangle]^{2}\cdot L)}$, where $L$ is the pathlength through the ionized gas, $\langle n_e\rangle$ is the average thermal electron density within the volume along the pathlength traced by the telescope beam, and $\langle B_{t,\|}\rangle$ is the strength of the average turbulent magnetic field along the LOS. We estimated the thermal electron densities through the emission measures: 
\begin{equation}
\label{EM_f-la}
EM = {\int_0}^{L}{n_e}^2 (l)dl  = \langle {n_e}^2 \rangle \cdot L = \frac{\langle {n_e} \rangle^2}{f_e} \cdot L
\end{equation}
where $f_{e}$ is the volume filling factor of the thermal electrons. The emission measure was computed following \citet{Valls-Gabaud1998} using the extinction-corrected H$\alpha$ emission (refer to Sect. 3.1.2 in  \citealt{Mora-Partiarroyo2019_A}). Thermal electron densities in NGC~4631 were then estimated assuming a volume filling factor of $f_{e} = 5-20\%$, according to observations in the Milky Way \citep{Berkhuijsen2006, Berkhuijsen2008}. With the observed Faraday depolarization distribution between data at 4.85 and 8.35~GHz derived by \cite{Mora2013}, we obtained the mean Faraday depolarization (DP$_{\rm{obs}}$(4.85, 8.35)) in the central region, the disk (excluding the central area), and in the disk-halo interface of NGC~4631.

The estimated $d/f$ ratios in different regions of NGC\,4631 are listed in Table~\ref{table_ratio}. The values represent upper limits since they were derived assuming that internal Faraday dispersion is the only effect responsible for the observed Faraday depolarization in NGC~4631. 
The ratio ranges between 4 to 15~pc for the central region of the galaxy. If we assume a filling factor of $f = 0.1$, the size of the turbulent cells in the central area of NGC~4631 is approximately $d\; \rm{\approx40-150}$~pc. For the disk-halo interface, if one takes into consideration the filling factor of the cells derived for the Milky Way \citep{Sun2009}, $f = 0.5$, we obtain a turbulent cell size of $d\; \rm{\approx92-320}$~pc. These values agree with those predicted by \cite{Sokoloff} for the disk and halo of spiral galaxies. In M\,51, \cite{Shneider2014} estimated that the size of the turbulent cells in the disk range between 40-61~pc and in the halo 157--253~pc, similar to our results.

\begin{table}[h]
\centering
\caption[Derived estimates for $\sigma_{RM}$ and the ratio d/f]{Observed Faraday depolarization between 4.85 and 8.35~GHz (DP$_{\rm{obs}}$(4.85,8.35)), emission measures (EM), pathlength (L) through the disk and halo towards the indicated regions, average thermal electron densities ($\langle n_e \rangle$), and turbulent magnetic field strength along the line-of-sight (B$_{t, \|}$) in different regions of NGC~4631. Estimates for the RM dispersion ($\sigma_{RM}$) and the ratio between the size of the turbulent cells and the filling factor of the cells ($d/f$) are given in the last two rows. }
\label{table_ratio}
\begin{minipage}{1.\columnwidth}
\begin{tabularx}{1\columnwidth}{c*4{>{\centering\let\\=\tabularnewline}X}}
\hline \hline
 & Central region & Disk\tablefootmark{a}& Disk-halo interface  \\ \hline
\rm{DP$_{\rm{obs}}$(4.85,8.35)}\tablefootmark{b}  & 0.4  & 0.5 & 0.8 \\ 
\rm{EM~[$\rm{ pc \cdot cm^{-6}}$]} & 2220 & 480 & 100\\
\rm{L~[kpc]}\tablefootmark{c} & 29 & 24 & 20\\ 
\rm{$\langle \rm n_e \rangle$~[$\rm{cm^{-3}}$]}& 0.06-0.12 & 0.03-0.06 & 0.016-0.03 \\ 
\rm{B$_{\rm t, \|}$~[$ \mu G$]}                      &9.3 &6.4 & 5.7 \\
\rm{$\rm \sigma_{\rm RM}$~[$\rm{rad / m^{2}}$]}& 300 & 255 & 132 \\
$d/f~\rm{[pc]}$                                                & 4-15 & 28-112 & 46-160  \\
\hline
\end{tabularx}
\tablefoot{
\tablefoottext{a}{Excluding central region.}
\tablefoottext{b}{Values obtained from \cite{Mora2013}.}
\tablefoottext{c}{Values obtained from \cite{Mora-Partiarroyo2019_A}.}
}
\end{minipage}
\end{table}

An increasing size of the turbulent cells with increasing distance from a galaxy's midplane can be explained in various ways. \cite{Hummel1991} proposed that the mass of rising cells is conserved, so that the cell size varies with $n_\mathrm{e}^{-3}$. If the cells are generated by Alfv\'en waves, their sizes may be proportional to the Alfv\'en velocity that increases with height above the midplane. This is due to the fact that the magnetic energy density decreases more slowly than the thermal energy density of the warm + hot gas, as shown by the much larger extent of the synchrotron emission compared to the thermal radio emission
(cf. Figs.~5 and 6 in  \citealt{Mora-Partiarroyo2019_A}).
The turbulent cells may also be related to sound waves that propagate with a speed increasing from disk to halo, due to the increase in gas temperature.

\section{Discussion}
\label{Discussion}

We detected for the first time a smooth, large-scale Faraday depth pattern in the halo of an edge-on spiral galaxy, as described in Section~\ref{FDdistribution}, indicating a regular (coherent) magnetic field structure. The Faraday depth value is proportional to the regular magnetic field strength along the line-of-sight (LOS), i.e., to the parallel component of the regular magnetic field. A positive value of the Faraday depth implies that the LOS component of the regular magnetic field (B$_{\rm reg}$) points towards the observer, while a negative value implies that the LOS component of B$_{\rm reg}$ points away from the observer.

The Faraday depth pattern in the northern halo of NGC~4631 (Figure~\ref{RM_Cbad_20.5arcsec}) reveals a systematic change in sign from east to west, while the sign of Faraday depth is constant along six regions which are oriented almost vertical to the galactic plane and reach heights of about 2\,kpc. Only in the north-eastern halo the regions of constant sign turn towards a more horizontal orientation.

\begin{figure}[h]
\centering
\includegraphics[trim = 0mm 0mm 0mm 0mm, clip,width=1\columnwidth]{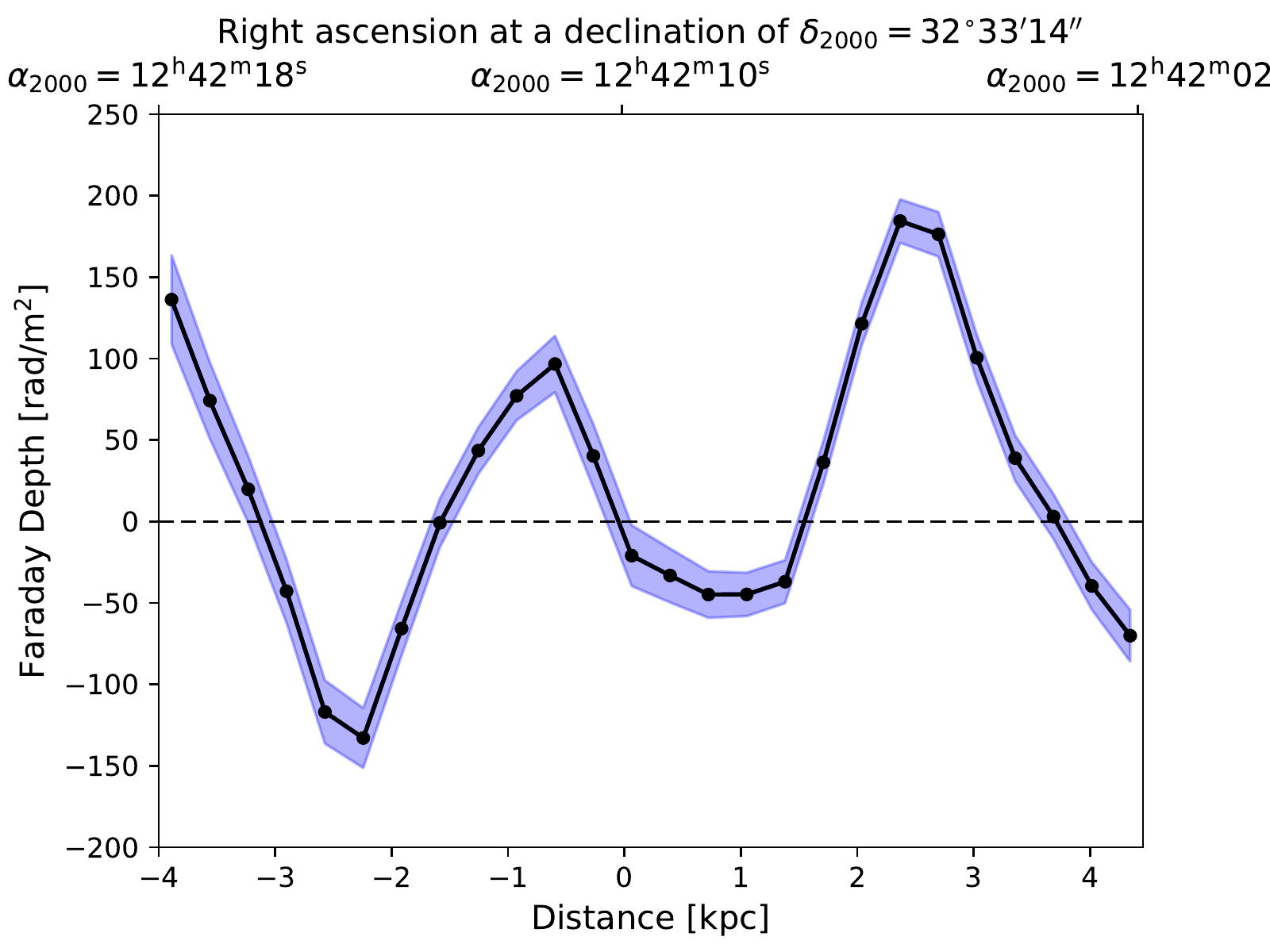}
\caption[Profile of the Faraday depth observed at C-band]{Profile of the Faraday depth distribution observed at C-band along a strip through the northern halo of NGC~4631 at a distance of about 1.7\,kpc from the galaxy's major axis and with a width of $20\farcs5$ (about 0.8\,kpc). The shaded blue region represents the error range. Distance = 0~kpc corresponds to the position $\alpha_{2000}=12^\mathrm{h}42^\mathrm{m}10^\mathrm{s}$, $\delta_{2000}= 32^{\circ}33'14''$. The magnetic field of NGC~4631 has a systematic change in direction on a scale of about 1.9\,kpc. The mean error in Faraday depth values is 16.5$~\rm{ rad / m^{2}}$}.
\label{profile}
\end{figure}

In Figure~\ref{profile} we present a profile
of the Faraday depth distribution in the northern halo of the galaxy extending from about $\alpha_{2000}=12^\mathrm{h}42^\mathrm{m}02^\mathrm{s}$ to $\alpha_{2000}=12^\mathrm{h}42^\mathrm{m}18^\mathrm{s}$ at a constant declination of $\delta_{2000}= 32^{\circ}33'14''$ (see blue rectangular box in Fig. \ref{RM_Cbad_20.5arcsec}). The Faraday depth varies smoothly from about $+200~\rm{rad / m^{2}}$ to about $-150~\rm{rad / m^{2}}$ in a quasi-periodic pattern. The Faraday depth distribution in the northern halo of the galaxy shows that the regular parallel magnetic field component has a systematic change in direction on a scale of about 1.9\,kpc (half wavelength). 

There are also regions with positive and negative Faraday depths in the southern halo (with one exception just south of the central region). However, the changes do not follow a regular pattern similar to the northern halo. The regions with constant sign of Faraday depth are horizontally orientated in the south-eastern part, in contrast to being orientated vertically in the north-eastern part.

A large-scale magnetic field has a parallel and a perpendicular component with respect to the LOS, $B_\parallel$ and $B_\perp$. The latter is traced by the orientation of the polarization angles of the synchrotron emission, which are presented in Figure~\ref{RM_Cbad_20.5arcsec} overlaid on the Faraday depth distribution. This magnetic field orientation exhibits an X-shaped pattern with strong vertical components above and below the central region. Some deviations from this pattern are visible in the eastern part of the halo, north and south of the major axis. 

\begin{figure}[h]
\centering
\includegraphics[trim = 0mm 0mm 0mm 0mm, clip,width=1\columnwidth]{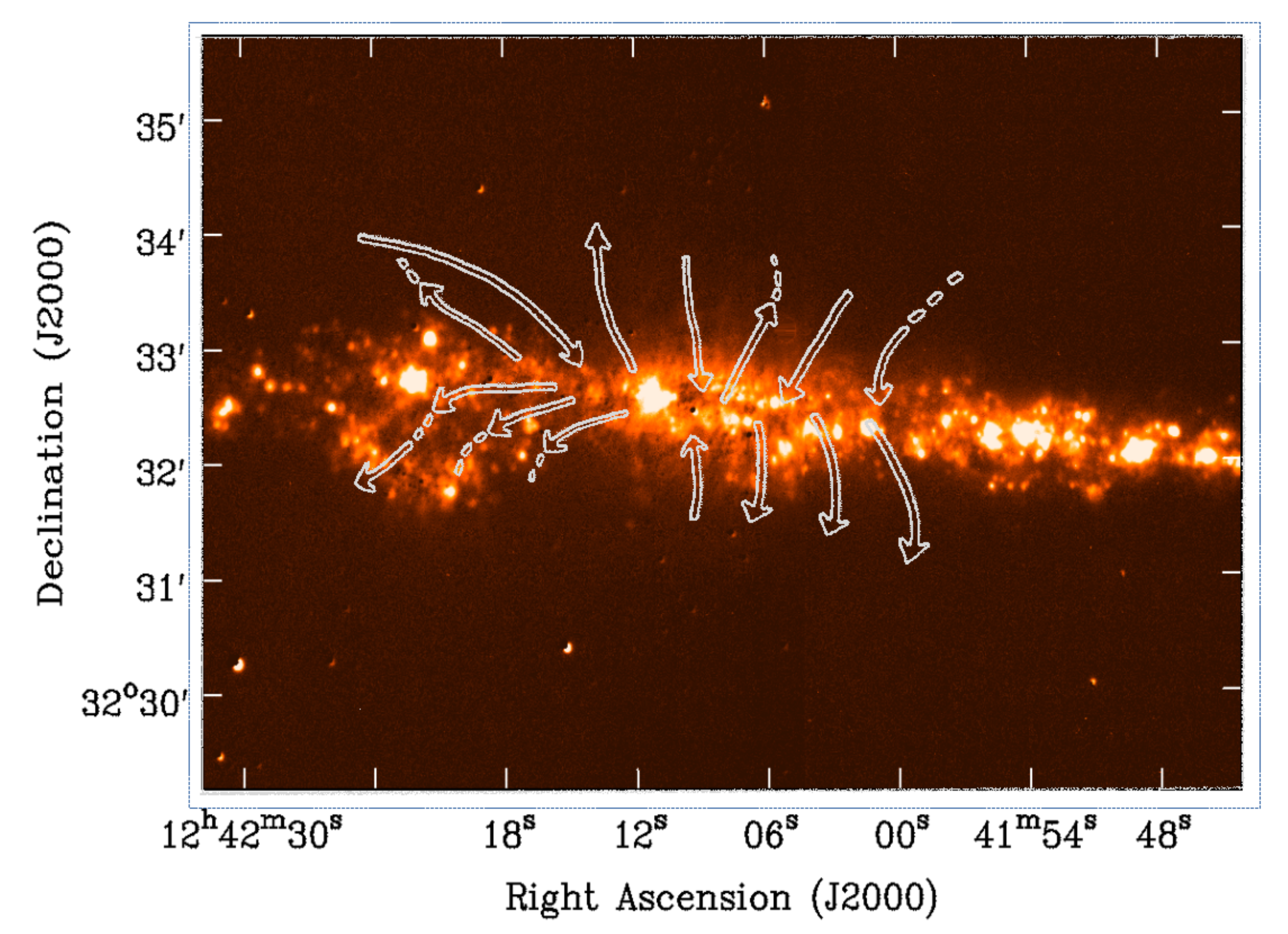}
\caption[Ropes]{Sketch of the giant magnetic ropes with alternating directions in the northern halo of NGC~4631 and the large-scale field in the southern halo. Dotted lines indicate that the field bends away from the observer, as evidenced by large-scale field reversals roughly parallel to the disk. The background H$\alpha$ image is from SINGS \citep{2003SINGS}}.
\label{ropes}
\end{figure}

Remarkably, the orientations of $B_\perp$ are, to a first order, roughly parallel to the regions of constant sign of Faraday depth, also in the north-eastern part of NGC~4631 where two such regions bend from a vertical orientation with respect to the disk towards a more horizontal orientation. A combination of $B_\perp$ with $B_{\mathrm{reg},\parallel}$ observed in Faraday depth indicates several large-scale magnetic field structures in the halo, extending perpendicular to the disk, which we call ``giant magnetic ropes (GMRs)'' (Fig.~\ref{ropes}).

The Faraday depth pattern in the south-eastern and north-western regions of the halo is different in the way that the reversals occur along lines that are not perpendicular, but roughly parallel to the disk,
which indicates that a large-scale magnetic field extends from the disk into the halo and bends away from the observer (Fig.~\ref{ropes}).

Our data allow us to describe the giant magnetic ropes in the northern halo in more detail. The high values of Faraday depth demonstrate that the field in the ropes must be inclined out of the sky plane, in order to get a component along the LOS. As the Faraday spectrum contains only one unresolved component, Faraday depth $FD$ is given as:
\begin{equation}
FD = 0.812 \, \langle n_\mathrm{e} \rangle \, \langle B_\mathrm{reg,\parallel} \rangle \, L \, ,
\end{equation}
where $\langle n_\mathrm{e} \rangle$ is the average electron density (in cm$^{-3}$), $L$ is the pathlength along the LOS (in pc), and $\langle B_\mathrm{reg,\parallel} \rangle$ (in $\mu$G) is the average strength of the regular field along the LOS. For $FD \approx \pm 150$\,rad/m$^{2}$ and $\langle n_\mathrm{e} \rangle \approx 0.02$\,cm$^{-3}$ (Table~\ref{table_ratio}), assuming that the thickness of a field rope is the same in all directions and that the rope is inclined out of the sky plane by the angle $\beta$, so that $L\approx 2\,$kpc\,/\,$\cos\, \beta$, we need a field strength of the regular field along LOS of $\langle B_\mathrm{reg,\parallel} \rangle \approx \pm 4.6\,\mu$G$\,\cos\, \beta$ to account for the observed $FD$. This field strength is an upper limit because the magnetic rope is probably associated with enhanced H$\alpha$ emission, so that the effective pathlength $L$ is smaller than 20\,kpc as assumed in Table~\ref{table_ratio}. For example, $\langle n_\mathrm{e} \rangle \approx 0.04$\,cm$^{-3}$ would give $\langle B_\mathrm{reg,\parallel} \rangle \approx \pm 2.4\,\mu$G$\,\cos \, \beta$.

The average strength of the ordered halo field perpendicular to the LOS derived from the synchrotron intensity and the degree of polarization (assuming energy equipartition) with a pathlength of 2\,kpc and a synchrotron spectral index of -1.0, results in $\langle B_{\perp} \rangle \approx 7\,\mu$G$\,(\cos \,\beta)^{1/4}$. This is an upper limit for the average strength of the regular field $\langle B_\mathrm{reg,\perp}\rangle$ perpendicular to the LOS because some fraction of the ordered field may be due to anisotropically turbulent fields that do not contribute to $FD$.

The effective inclination of the magnetic rope is:
\begin{equation}
\tan\, \beta = \langle B_\mathrm{reg,\parallel} \rangle /\langle B_\mathrm{reg,\perp} \rangle \, ,
\end{equation}
which yields $\beta \geq 30\degr$ for $\langle n_\mathrm{e} \rangle \approx 0.02$\,cm$^{-3}$ or $\beta \geq 18\degr$ for $\langle n_\mathrm{e} \rangle \approx 0.04$\,cm$^{-3}$. Assuming $\beta \approx 50\degr$ for the first case and $\beta \approx 30\degr$ for the second case gives $\langle B_\mathrm{reg,\parallel} \rangle \approx \pm 3.0\,\mu$G and $\approx \pm 2.1\,\mu$G. Then the strength of the regular field is $|B_\mathrm{reg}| = |B_\mathrm{reg,\parallel}| / \sin \, \beta \approx 4\,\mu$G for both cases.

The field reversals can be explained in two ways: (a) The inclination changes sign, i.e., the ropes are inclined alternatively towards us and away from us. However, we are not aware of a mechanism that is able to reverse the radial field component on scales of a few kpc. (b) A more plausible configuration is that the magnetic rope is oriented perpendicular to the plane, but has a conical shape. The field component along the LOS changes sign from the far side to the near side. The observed average $FD$ is that of the near side, because Faraday rotation of polarized emission from the far side is canceled when passing through the near side.
The field forms conical ropes with significant alternating directions. The conical shape means that the field has significant radial components. A large radii, the radial components become visible as tilts of the field lines in the sky plane (see Fig.~\ref{Cband_POLI_7arcsec}).

On the whole, the observed pattern in the halo of NGC~4631 does not reflect a simple large-scale dipolar or quadrupolar magnetic field in the halo as expected from the thin-disk approximation of the mean-field dynamo theory for galaxy disks \citep[e.g.,][]{Ruzmaikin+1988}. This, however, is not surprising as the observed magnetic field strength in the halo is comparable to the disk field strength, while in those models the dipolar or quadrupolar halo field that accompanies the plane-parallel disk field is predicted to be about a factor of 10 weaker than the disk field. The observed pattern also does not fit to the analytical models of the X-shaped fields in galactic halos by \cite{Ferriere+2014}. Numerical simulations, as e.g., the galactic-scale MHD simulations of a CR-driven dynamo \citep{Hanasz2009}, indicate a more patchy pattern.

The scale-invariant mean-field dynamo theory \citep{Henriksen2017,Henriksen+2018} predicts both axially symmetric and non-axially symmetric magnetic spiral arms rising into the halo. Such a theory differs from disk-based dynamo theory by allowing some dynamo activity to continue into the galactic halo. The spiral flux tubes of both types are wound on conical surfaces which, at the equator, are nearly parallel to the equator, but close up around the minor axis. The axially symmetric \& conically symmetric spirals project into axially symmetric magnetic arms (ASS). The non-axially symmetric \& conically symmetric spirals project into bisymmetric magnetic arms (BSS) in a face-on view of a galaxy. In this theory, the X-type field behavior is associated with the axially symmetric fields. The components of the non-axially symmetric field parallel to the disk are predicted to be anti-symmetric on crossing the disk, while the parallel components of the axially symmetric field may have either even or odd symmetry across the disk. Strong non-axially symmetric components produce field reversals. The signs of the non-axially symmetric field along the line-of-sight alternate with radius, where the number of reversals depends on the mode number. They also contain field lines looping over and under the magnetic arms (which are therefore themselves helices) rather than diverging from the disk in an X-type fashion. Dynamo models including outflows provide a reasonable fit to the magnetic field structure and the reversals in the northern halo of NGC~4631, while the best results are found for accretion models \citep{woodfinden19}.

The observations presented in this paper are basically consistent with conically symmetric \& non-axially symmetric fields. Improved observations are needed in future. Detailed modeling of the magnetic field in the halo of NGC~4631, taking into account the polarization and depolarization properties of the galaxy, is necessary and will give more insight into the origin of these features.

\section{Summary and conclusions}
\label{Conclusions}

Radio continuum observations of NGC~4631 were performed with the Karl G. Jansky Very Large Array at C-band (5.99~GHz) in the C \& D array configurations, and at L-band (1.57~GHz) in the B, C, \& D array configurations. Our conclusions are as follow:

\begin{itemize}

\item We detected linearly polarized emission at C-band and L-band.
The application of RM-synthesis to the data allowed us to map the Faraday depth at C-band and L-band.

\item The magnetic field in the halo is characterized by strong vertical components above and below the central region of the galaxy. The magnetic field in the disk is only clearly seen in the eastern side of NGC~4631, where it is parallel to the plane of the major axis of the galaxy. The nonthermal degree of polarization at C-band (L-band) is within a few percent in the disk and increases to 20--30\% (35--45\%) in the halo.

\item We detected, for the first time, a large-scale Faraday depth pattern in the halo of an external spiral galaxy, which implies the existence of a regular (coherent) magnetic field in the halo with a strength of $\approx 4\,\mu$G. 
The observed Faraday depth pattern shows several large-scale field reversals in the northern halo.
To our knowledge, \textit{this is the first time such reversals have been seen in an external galaxy.} As an explanation, we propose giant magnetic ropes (GMRs) of conical shape, which are oriented perpendicular to the galactic disk and have alternating directions.
\end{itemize}

\begin{acknowledgements}
We thank Dominic Schnitzeler for fruitful discussions and feedback in the application of the RM-synthesis algorithm. AB acknowledges financial support by the German Federal Ministry of Education and Research (BMBF) under grant 05A17PB1 (Verbundprojekt D-MeerKAT). RAMW acknowledges support from NSF Grant AST-1615594.

\end{acknowledgements}

\bibliographystyle{aa}
\footnotesize
\bibliography{References_NGC4631_AA_2017} 
\end{document}